\documentclass[aps,prb,groupedaddress,twocolumn,10pt]{revtex4-1}

\begin{document}

\newcommand{\be}{\begin{equation}}
\newcommand{\ee}{\end{equation}}
\newcommand{\bea}{\begin{eqnarray}}
\newcommand{\eea}{\end{eqnarray}}

\title{Phase space holography with no strings attached}

\author{D. V. Khveshchenko}
\affiliation{Department of Physics and Astronomy, 
University of North Carolina, Chapel Hill, NC 27599}

\begin{abstract}
\noindent
This note discusses the Wigner function representation from the standpoint of establishing a   
holography-like correspondence between the descriptions of a generic quantum system 
in the phase space ('bulk') picture versus its spacetime ('boundary') counterpart.  
Under certain circumstances the former might reduce to the classical dynamics of a local 
metric-like variable while the latter takes on the form of some bosonized collective 
field hydrodynamics. This generic pseudo-holographic duality neither relies on any 
particular symmetry of the system in question, nor does it require any connection to an
underlying 'string theory', as in the various 'ad hoc' scenarios of applied holography.  

\noindent
\end{abstract}
\maketitle

{\it Generalized holography}\\

Among all the remarkable theoretical advances under the hashtag ${\#}holography$ 
the one of the greatest interest to condensed matter physics is the ongoing quest for generalizing the original, highly constrained (i.e., Lorentz- and maximally 
super-symmetric, etc.), 
string-theoretical holographic conjecture to as broad as possible 
(i.e., non-Lorentz-, non-isotropic-, non-super-symmetric, etc.) variety of quantum many-body systems \cite{ads}.  
 
In that regard, the so-called 'bottom-up' holography has been portraying itself as a powerful technique offering solutions to the traditionally hard condensed matter problems. 
This innovative approach (sometimes referred to as the anti-de-Sitter/condensed matter theory, or $AdS/CMT$, correspondence) 
borrows its formal structure and mathematical apparatus (often 'ad verbatim', for the lack of an alternative) from the very specific, carefully crafted, and highly symmetric examples of such duality known under the acronym $AdS/CFT$ and conjectured in the original context of fundamental string theory and its various reductions \cite{ads}. 

So far, however, all the attempts of putting applied holography on a firm foundation 
- either along the lines of the geometrized renormalization group (RG) 
flow or entanglement dynamics 
in tensor networks, or by using artificial thermodynamic (Fisher-Ruppeiner, Fubini-Study, etc.) metrics, or else - have remained consistently inconclusive. 

Nonetheless, instead of striving to deliver a solid proof of principle, 
the field of $AdS/CMT$ has, by and large, stayed
the course reminiscent of hacker's code cracking: that is, trying to guess some 
higher-dimensional enhanced gravity-like theory 
(often on the sole basis of technical convenience) 
and then rely on the persuasive power of visual agreement between 
some pre-selected experimental data plots and the (for the most part, numerical) calculations 
based on the above $AdS/CMT$ 'dictionary'. A great many number of the customarily verbose and look-alike accounts of such pursuits can be readily found in the applied holographic literature from the year $2007$ onwards \cite{holo}.  
Judging by the factual outcome of this massive attack, though, the code does not appear to have been cracked yet.

Such fundamental shortcomings notwithstanding, 
the still continuing occasional exercises in the holographic phenomenology  
utilize a handful of the popular bulk geometries, especially focusing 
on the legacy solutions of the prototypical Einstein-Maxwell-dilaton theory. 
Other than their relative simplicity and sheer availability 
(alongside, possibly, some lingering anthropic factor) 
a proper justification of such choices does not appear to have been an essential part of the holographic agenda, regardless of 
whether the results were meant to be applied to the lattice Hubbard-like models,
superconducting cuprates, low-density $2DEG$, Dirac/Weyl materials such as graphene, or else.       
 
Lately, though, a gradual decline in such 'orthodox' applications of the $AdS/CMT$ machinery has been giving way to advanced hydrodynamics of strongly coupled quantum matter and general
out-of-equilibrium physics (eigenstate thermalization, many-body (de)localization, 
chaos spreading, operator growth, etc.).
Correspondingly, instead of the once ubiquitous renditions of esoteric black holes, 
nowadays a slide show on the topic of $AdS/CMT$ is more likely to feature  
the images of water flows, rapids, whirlpools, and other familiar hydrodynamic patterns \cite{video}.  

Of course, hydrodynamics, while suggesting some intriguing holographic
connections, has long been discussed outside of any holographic context.
Therefore, the renewed appreciation and novel applications (thanks to a number 
of recent experimental advances) of 
the theory as Earthly as hydrodynamics alone do not provide 
an answer to the question that should have (but does not seem to have been) 
long dominated the holographic discourse, that is: 'So why, on Earth, strings?!'.

The goal of this note is to recall a decades-old theoretical approach 
known as collective field theory \cite{das} 
and its more recent developments that might be capable 
of providing a much-needed justification for the 'stringy hydrodynamics' (especially, in 
those non-relativistic and/or rotationally-non-invariant settings that are typical to the condensed matter applications 
but do not normally occur in the original string-theoretical context). 
Alternatively, this approach can be viewed as a variant of the long-pursued idea of 
(non-linear) 'bosonization' that aims to reformulate a quantum theory of interest in terms of some intrinsically geometric bosonic variables. 

Notably, while demonstrating some features reminiscent of the desired holographic
correspondence, this approach shows that the pertinent space-time metrics may not be chosen at will and can often be quite different from the routinely utilized ones.
\\

{\it Phase space quantization}
\\

A systematic description of many-body dynamics in $d$ spatial dimensions calls for the use of the Wigner function $w({\bf x},{\bf p},t)$ defined in the $2d+1$-dimensional phase space (plus time). The space of such functions is equipped with the Moyal product 
\be 
f({\bf x},{\bf p})\star g({\bf x},{\bf p})
=
f({\bf x},{\bf p})e^{i{\hbar\over 2}({\overleftarrow \partial_{\bf x}}{\overrightarrow 
\partial_{\bf p}}-{\overleftarrow \partial_{\bf p}}{\overrightarrow \partial_{\bf x}})}
g({\bf x},{\bf p})
\ee
which introduces the symplectic structure through the Moyal bracket 
\be
{\{}f, g{\}}_{MB}=f\star g  - g\star f
\ee
The Wigner's description is well suited for taking into account the  
underlying theory's invariance under the phase space volume-preserving diffeomorphisms,
including its natural time evolution that can be thought of as  
a canonical transformation governed by the Liouville theorem. 

Furthermore, when quantizing the system via the method of functional integration, 
the function $w({\bf x},{\bf p},t)$ becomes a 
constrained field variable implementing a coadjoint orbit's quantization 
of the phase space volume-preserving diffeomorphisms a la Kirillov-Kostant. 

In this procedure, an orbit's element ${\hat g}|\Psi_0><\Psi_0|{\hat g}^{-1}$ is 
constructed by acting with an element  
$\hat g$ of the infinite-dimensional group of diffeomorphisms on 
the projector to a chosen (e.g., ground) state $|\Psi_0>$. 
In the partition function 
$
Z=\int Dw \exp(-S(w))
$
the integration then runs over the functions satisfying the constraints
\be
w
\star w=w~~~~~~~Tr w=1
\ee
and governed by the action 
\be
S(w)=\int dxdpdt (i\int^1_0ds w{\{}\partial_{\tau}w,\partial_{s}w{\}}_{MB} - wH)
\ee
where the first term represents the Berry phase where the integral over the auxiliary variable $s$ depends only on the 
boundary value $w({\bf x},{\bf p},t,s=1)=w({\bf x},{\bf p},t)$ at $s=1$. 
This way, one arrives at the formally exact geometrized
description of the non-linear $\sigma$-model type \cite{das}.  

The equation of motion derived from Eq.(4) 
\be
{\dot w}+{\{}w,H{\}}_{MB}=0
\ee
reproduces the standard kinetic equation when the Moyal bracket is approximated,
to lowest order in the powers of $\hbar$, by the Poisson one 
(hereafter the dot and prime stand for the time and space derivatives, respectively)
\be
{\dot w}+w^{\prime}\partial_{\bf p}H_1-H_1^{\prime}\partial_{\bf p}w=St[w]
\ee
where the one-body Hamiltonian $H_1$ may include an external potential. For example, in the so-called 'non-critical $2d$ string theory'  
where the spatial coordinate originates from the eigenvalues of $N\times N$ 
matrices it happens to be the inverted oscillator ($V\sim -x^2$) \cite{das}.  
Also, the $n\geq 2$-body terms $H_n$ in the Hamiltonian involving higher powers of $w$ are bundled into the collision integral in the right hand side.

In the case of fermions, the semiclassical vacuum configuration corresponding to the uniform Fermi sea is described by the expression
\be
w_0({\bf x},{\bf p},t)=\theta(\mu({\bf x},t)-\epsilon_{\bf p})
\ee
where the local chemical potential $\mu({\bf x},t)$ denotes a sharp 
boundary between the occupied and vacant momentum 
eigenstates with the dispersion $\epsilon_{\bf p}$.  

Facilitating further progress with the $2d+1$-dimensional 
'bulk' theory (4) requires a convenient parametrization of the bounding momentum. 
Previously, a similar task was tackled in the early works on multi-dimensional bosonization where this goal was achieved by distinguishing 
between the Fermi momentum $\bf p_F$ tracing 
the fiducial Fermi surface (FS) and the normal to the FS ('radial') degree of freedom
describing fluctuations of the momentum distribution \cite{bos}. 
 
For instance, in the much studied $d=2$ case the simplified action for  
the vector $\bf k_F$ reads  
\be
S({\bf p_F})=
\int dx dt (i\int^1_0  ds  {\bf p_F} \partial_{\tau}{\bf p_F}
\times
\partial_{s}{\bf p_F} -H({\bf p_F}))
\ee 
where the Hamiltonian $H$ is cast in terms  
of the local density $\rho={\bf p_F}\times\partial_t{\bf p_F}/2$.  

In what follows, we focus on the case of $d=1$ where the fluctuating FS  
can be described in terms of $M\geq 1$ pairs of the Fermi momenta $p^{(n)}_{\pm}(x,t)$
bounding the occupied states ($M>1$ accounts for the possibility of 'folds' in the presence of 
shock waves and other FS singularities \cite{das}).  

In particular, the $1d$ configuration (7) reads     
\be
w_0(x,p,t)=\sum_{\pm}\sum_{\alpha=1}^M
(\pm)\theta(p^{(\alpha)}_{\pm}(x,t)-p)
\ee
while its small perturbation  
\be
\delta w(x,p,t)={\hbar}\delta(x-x_{cl})\delta(p-p_{cl})
\ee
is strongly peaked
at the classical phase space trajectory ($x_{cl}(x_0,p_0,t)$, $p_{cl}(x_0,p_0,t)$) 
where the initial data $x_0$ and $p_0$ are determined 
by the current values ($x, p$) at a later time $t$. 

It is worth mentioning, though, that while being capable of faithfully 
reproducing the long-distance, 
late-time asymptotics of the response functions, in its practical (hence, approximate) form 
the $d>1$-dimensional bosonization technique    
is not well suited for computing the Lindhard-type $2p_F$-singularities, just as it may 
not be sufficient for the single-particle propagators \cite{bos}.
 
In $d=1$, despite several decades of studies there has been a recent 
surge of renewed interest in the out-of-equilibrium dynamics of 
quantum interacting bosons and fermions. Many of those studies focus on the integrable and non-ergodic systems which are governed by the generalized Gibbs ensembles (GGE) and may not comply with the more generic eigenstate thermalization hypothesis (ETH) \cite{hydro}.    

It is worth noting that in $1d$  Eq.(2) represents a classical analog
of the infinite-dimensional quantum algebra $W_{\infty}$  
composed of the operators ${\hat W}_{mn}=
({\hat x})^m({\hat p})^n$ 
with the commutation relations 
\bea
[{\hat W}_{mn},{\hat W}_{rs}]=\sum_{k=1}{(-\hbar)^k\over k{\!}}~~~~~~~~~~\\
({n{!}r{!}\over (n-k){!}(r-k){!}}- {m{!}s{!}\over (m-k){!}(s-k){!}}) {\hat W}_{m+r-k,n+s-k}\nonumber
\eea
where the r.h.s. reduces to $(ms-nr){\hat W}_{m+r-1,n+s-1}$ in the limit
$\hbar\to 0$ limit, thereby encompassing the algebra $SL(2,R)$. 
This algebra has been extensively studied in the context of Quantum Hall Effect (QHE)  
and the various reincarnations of (effectively) non-commutative spacetimes. 

An abstract Hilbert space can be readily equipped with a geometric structure 
that has long been elucidated alongside 
the more familiar Berry phase. However, the even (Fubini-Study metric) 
- as opposed to the odd (Berry curvature) - component
of the same rank-$2$ tensor has been receiving less attention.  

Such a phase space metric can be naturally introduced in 
the context of special coherent  
- (de)localized neither in the coordinate, nor momentum space - states  
$|p,x,0>=e^{i{\hat P}x-i{\hat Q}p}|0>$ which 
minimize, both, the coordinate and momentum uncertainties.

Allowing, for the sake of generality, some coordinate-momentum 
cross-correlations the corresponding Wigner function reads  
\bea
w_{coh}(x,p,0)=\int dy e^{ipy}<\Psi|x+y/2><x-y/2|\Psi>\nonumber\\
={\hbar\over D^{1/2}}\exp(-{\sigma_p\delta x^2+
\sigma_x\delta p^2+2\sigma_{xp}\delta x\delta p\over {2D}})~~~~~~~~~~~
\eea
where $\delta x=x-x_0$, $\delta p=p-p_0$, $D=\sigma_x\sigma_p-\sigma^2_{xp}$, and  
 the parameters $\sigma_x,\sigma_p,\sigma_{xp}$ control the Gaussian 
 coordinate and/or momentum variations. 

The above suggests a naturally defined Fubini-type metric on the phase space \cite{qm}
\be
ds^2=(<\partial_{\mu}\Psi|\partial_{\nu}\Psi>-
<\Psi|\partial_{\mu}\Psi><\partial_{\nu}\Psi|\Psi>)\delta_{\mu}\delta_{\nu}
\ee
where $\delta_{\mu}=(dx,dp)$. 
Taken at its face value, this formula establishes some form of superficial correspondence 
between single-particle quantum mechanics and 
$2d$ metrics that can be viewed as solutions of certain classical gravity. 

Further generalizing Eq.(13) to include energy fluctuations one arrives at the (Euclidean) $3d$ metric 
\bea
ds^2=<(\Delta {\hat H})^2>dt^2+<(\Delta {\hat x})^2>dp^2~~~~~~~~~~\nonumber\\
+<(\Delta {\hat p})^2>dx^2+2<\Delta {\hat x}\Delta {\hat p}>dxdp~~~~~~~~~~
\eea
given by the uncertainties of 
the conjugate variables ($x\leftrightarrow p$, $t\leftrightarrow H$).
Also, considering the metric (14) to be the expectation value $ds^2=<\Psi|d{\hat s}^2|\Psi>$
of the operator-valued interval $d{\hat s}^2$ paves the way for promoting the 
bulk (phase space)-to-boundary (spacetime) relationship to the quantum level. 

As the operators' uncertainties depend on the choice of the state $|\Psi>$, 
so does the dual metric (14). Heuristically, one might expect that
for the single-particle dispersion governed by the dynamical exponent $z$
($\epsilon_{\bf p}\sim {\bf p}^z$)  the above variances behave as follows
\bea
<(\Delta p)^2>\sim <(\Delta x)^2>^{-1}\sim p^2\nonumber\\
<(\Delta x)(\Delta p)>\sim <(\Delta p)>^{1/2}<(\Delta p)^{-1/2}\sim p^0\nonumber\\
<(\Delta H)^2>\sim <(\Delta p)^2>^z\sim p^{2z}
\eea
so that the metric (14) conforms to the so-called Lifshitz variety  
(the coefficients $A,B,C,D$ are constants)
\be
ds^2=Ap^{2z}dt^2+Bp^2dx^2+C{dp^2\over p^2}+2Ddxdp
\ee
which has been often invoked in the various applications of $AdS/CMT$ \cite{ads}.
\\

{\it Non-linear hydrodynamics}
\\

The formally exact representation (4) of the phase space dynamics
provides a basis for further simplifications, thus giving rise to (semi)classical hydrodynamic equations for the various moments of the Wigner distribution
\be
w_n({\bf x},t)=\int d{\bf p} w({\bf x},{\bf p},t){\bf p}^n
\ee
Among those moments are such standard hydrodynamic variables
as the local mass $\rho$ ($n=0$), momentum $Q$ ($n=1$), and energy $\epsilon$ ($n=z$)  
densities, respectively. 
This transition from the entire Wigner function to
the first few of its moments can be thought of as a dimensional reduction from the $2d+1$-dimensional bulk (phase space) to its $d+1$-dimensional boundary hypersurface (spacetime)
which, in practice, amounts to mere integration over the $d$-dimensional momentum.

Correspondingly, the mass $J_{\rho}$, momentum $J_{Q}$, 
and energy $J_{\epsilon}$ currents are given by the general expression
\be
{\bf J}_{n}=\int d{\bf p}{\partial\epsilon_p\over \partial {\bf p}}w({\bf x},{\bf p},t){\bf p}^n 
\ee 
for $n=0,1,2$. 
 
In the case of $d=1$ the lowest moments of the Wigner function 
(17) correspond to the aforementioned bounding Fermi 
momenta $p_{\pm}=\int dp w (1\pm sgn p)/2$  or, equivalently, the local density and material velocity 
\be
\rho= {1\over 2\pi}(p_{+}-p_{-})~~~~~~ v= {1\over 2m}(p_{+}+p_{-})
\ee
Note that limiting the momentum values to the interval 
$0<p<\infty$, similar to the holographic radial variable, is dictated by the
 chiral nature of the excitations carrying sign-definite momenta.  

These variables have the Poisson bracket  
\be
{\{}\rho(x),v(y){\}}=\partial_x\delta(x-y)
\ee 
Then taking the various moments of Eq.(5) one arrives at the hydrodynamic equations  
of motion which include the continuity equation 
\be
{\dot \rho}+(\rho v)^{\prime}=0
\ee  
and the inviscid Navier-Stokes (a.k.a. Euler/Burgers/Hopf) one 
\be
{\dot v}+vv^{\prime}=-{P^{\prime}\over \rho} - \kappa{V^{\prime}\over \rho}
\ee
where the 'quantum pressure' $P(\rho)$ (internal stress tensor of the 
$1d$ quantum fluid) is a system-specific 
function of the local density, while the last term with
the dispersion curvature $\kappa={\partial^2_p\epsilon_p}$ represents the 
force exerted by the external potential (if any). 

It is well known that the non-linear hydrodynamic equations (21,22) can be derived even from
the free particle Schroedinger equation \cite{qm}.   
Specifically, by applying the Madelung 
parametrization of the wave function
$\Psi(x,t)={\sqrt \rho(x,t)}e^{iS(x,t)}$ and separating out the real and 
imaginary parts one arrives at 
the coupled continuity and Navier-Stokes equations, respectively,
where
\be  
v={1\over 2im}({\Psi^{*}\over \Psi})^{\prime}~~~~~~\rho=|\Psi|^2
\ee
In the r.h.s. of (22) the pressure 
\be
P={\hbar^2\over 8m}{(\rho^{\prime})^2\over \rho}
\ee
contributes towards the overall 
energy density $\epsilon=\rho v^2/2+P$ which, 
in general, might be neither 
polynomial, nor separable as a sum of two chiral components $P_{\pm}(p_{\pm})$. 

It appears, however, that the pressure gradient term couples excitations with opposite chiralities (left/right moving) at the level of operators with dimensions
of four or greater. Moreover, even if present, the non-chiral 
corrections do not affect the states which are composed exclusively of   
the chiral excitations with $p_+=0$ or $p_-=0$. 

Thus, an arbitrary single-particle state 
$
\Psi(x,t)=(2\pi i\hbar t/m)^{-1/2}\int dye^{im(x-y)^2/2\hbar t}\Psi_0(y)
$ 
of the free Hamiltonian of mass $m$
with the initial data $\Psi_0(x)$ 
provides a valid solution to the hydrodynamic equations (21,22) with the pressure (24).
Correspondingly, the pair of functions $\rho(x,t)$ and $v(x,t)$ 
determines a certain dual metric, as explained below.  
\\

{\it Solvable hierarchies}
\\

For certain choices of the Hamiltonian $H$   
Eq.(22) appears to belong to the infinite KdV (Korteweg–de Vries) 
hierarchy of integrable $1d$ systems \cite{hydro}. Such  
Hamiltonians $H^{\pm}_k$ are related to the Gelfand-Dickey polynomials
and form an infinite set of integrals of motion in involution ($[H^{\pm}_n,H^{\pm}_m]=0$).
In the asymptotically free regime of large momenta (high energies) 
the  $k^{th}$ member of this family describes small-amplitude 
excitations with the dispersion exhibiting the dynamical exponent $z=2k-1$.

In particular, the generic $1d$ Luttinger liquid (LL) behavior is associated 
with the quadratic Hamiltonian $H=\sum_{\pm}H^{\pm}_1$ given by the standard Sugawara construction 
\be
H_1={1\over 2}
\sum_{\pm}p^2_{\pm}=
{1\over 2}(p_{+}^2+p_{-}^2)
\ee
which gives rise to the equation of motion 
\be
{\dot p_{\pm}}\mp p_{\mp}^{\prime}=0
\ee
whose solutions given by the (anti)holomorphic functions $p_{\pm}(x_{\pm})$
describe two decoupled chiral ($x_{\pm}=x\pm t$) pseudo-relativistic ($z=1$) excitations.

Corrections to the LL Hamiltonian (25) may come from, both, 
the Gaussian terms of higher operator dimensions  which represent non-linear terms in the  dispersion of the collective $\rho$- and $v$-modes,  
as well as from the non-Gaussian ones which are due to some intrinsic non-linearity of the 
$1d$ dispersion, as in the case 
of non-relativistic fermions at a finite density \cite{hydro}.

For example, the next ($2^{nd}$) member of the KdV family is given by the non-Gaussian expression
\bea
H_2=\sum_{\pm}
({\pm}) 
{1\over 3}p_{\pm}^3+{1\over 2}(p^{\prime}_{\pm})^2
=\nonumber\\
{1\over 2}\rho v^2 +{\pi^2\over 6}\rho^3+ {1\over 2}(v^{\prime})^2
+{\pi^2\over 2}(\rho^{\prime})^2
\eea
for which the chiral components of Eq.(22) still remain uncoupled  
\be
{\dot p_{\pm}}\pm 3p_{\pm}p_{\pm}^{\prime}+p_{\pm}^{\prime\prime\prime}=0
\ee
In the asymptotic regime of high energies the linearized    
Eq.(28) describes small waves with the expressly Lorentz-non-invariant cubic dispersion ($z=3$). 

In the opposite, low-energy and essentially non-perturbative, limit Eq.(28) permits non-linear solitonic excitations ('cnoidal' waves)  \cite{hydro} 
\be
v(x,t)\sim {1\over \cosh^2x_{\pm}}
\ee
whose propagation is described by the dispersion 
$\epsilon_p\sim p^{5/3}$. Compactifying the spatial 
coordinate into a finite-length circle 
would then replace (29) with the elliptic Jacobi function.  

In general, the non-Gaussian terms in the Hamiltonian are sensitive 
to the microscopic details of the model
and stem from, both, kinetic and potential terms in the total energy.
Specifically, in the case of hard-core bosons, such as the Tonks-Girardeau limit
of the Lieb-Liniger model,
the Hamiltonian includes the pressure term $P(\rho)\sim \rho^3$.
By contrast, in the quantum Toda chain the function $P(\rho)$ is non-polynomial. 
However, despite not being dividable into a 
sum of two chiral terms, the latter can still fit into the KdV Hamiltonian (27) 
as the non-chiral terms appear to be irrelevant at low momenta  \cite{hydro}. 

Likewise, the deviations from the LL regime associated with a finite dispersion 
curvature and/or chiral interactions can be studied with the use of a linear 
combination ${\hat H}_1+{\hat H}_2+\dots$ of Eqs.(25) and (27) \cite{}. 
This way, one can obtain non-linear corrections $\delta\epsilon_p\sim p^{5/3}$ 
to the linear LL spectrum at small momenta \cite{hydro}. 

Moreover, Eq.(27) can be further modified by including irrelevant 
non-Gaussian terms, such as  $p^4_{\pm}$, without destroying its integrability.   
Indeed, such extension results in yet another, Gardner, equation (a.k.a. mixed KdV-m(modified)KdV, which two equations are related by virtue of the Miura transformation $p_{\pm}\to p^2_{\pm}+p_{\pm}^{\prime}$).

As a hallmark of integrability, the higher-$k$ level members of the KdV hierarchy 
possess the bi-Hamiltonian structure relating them as follows \cite{sh}
\be
\partial_x{\delta H^{\pm}_{k+1}\over \delta p_{\pm}}=
{\cal D}^{\pm}_x{\delta H^{\pm}_{k}\over \delta p_{\pm}}
\ee 
where the long derivative is 
\be
{\cal D}^{\pm}_x=2p_{\pm}\partial_x +\partial_x p_{\pm}+\partial^3_x
\ee
The higher level-$k$ members of the KdV and mKdV families 
can also be morphed into a two-parameter Gardner sequence of Hamiltonians. 
Furthermore, certain solvable systems of $M\geq$ coupled non-linear equations were shown 
to be associated with the higher-spin symmetry algebras $SL(M,R)$ (e.g., the Boussinesq equations for $M=3$) \cite{hs}. 

Generic Hamiltonians   
$
\sum_k\mu_k{H}^{\pm}_k
$
which includes different members of the integrable family can be used to 
derive zero entropy GGE hydrodynamics. In particular, one can study crossovers between the 
LL and higher level-$k$ regimes at varying momenta. 
In essence, this construction provides a (formally exact) 
bosonization scheme that was fully exploited, e.g., in the context 
of the solvable Calogero-Sutherland model \cite{hydro}. 

Under the time evolution governed by a superposition of the different   
$H_n$ a generic initial condition produces a collection of solitons 
with different velocities and a continuum of decaying dispersive waves.
Being more robust the soliton excitations dominate in the late-time behavior 
and, in particular, the system's  equilibration towards 
a steady GGE state described by the density matrix 
${\hat {\cal G}}=\exp(-\sum_k\mu^{\pm}_k{\hat H}^{\pm}_k)$ where the 
chemical potentials $\mu_k$ are to be determined 
by equating the averages of the commuting charges 
$<{\hat H}^{\pm}_k>=Tr({\hat {\cal G}}{\hat H}^{\pm}_k)/tr {\hat {\cal G}}$ to their chosen values. 

Along these lines, one can also 
study the von Neumann entropy $S=-Tr {\hat {\cal G}}\ln {\hat {\cal G}}$.
The Wigner function satisfying the constraints (3) 
corresponds to a pure state of zero entropy and the presence of an infinite 
number of conserved charges precludes standard thermalization.
When the constraint ceases to hold the state becomes mixed, thus resulting  
in a finite entropy. The ensuing thermalization can be accounted for 
by introducing viscous terms, such as $\eta \rho^{\prime\prime}$, in the r.h.s. of Eq.(22).
\\

{\it Dual bulk gravity}
\\

A deep relationship between classical gravity and hydrodynamics has 
long been known as one particular take on the holographic paradigm, 
often referred to as the 'fluid-gravity' correspondence.  
The crux of the matter is observation of the similarity between the asymptotic near-boundary behavior of the Einstein equations for the bulk metric and the Navier-Stokes ones describing a dual boundary 
fluid in one lesser dimension (besides, the complementary hydrodynamic 
behavior near the event horizon can be similar to that at the boundary).
Albeit being truncated and, therefore, approximate such relations can be 
systematically improved, thus enabling certain computational simplifications.    
Whether or not this duality can be promoted to the quantum level requires further analysis. 

Remarkably, in the case of $d=1$ this correspondence becomes exact.  
Specifically, the Einstein equations stemming from the action 
of $3d$ gravity with a negative cosmological constant 
(here $l$ and $G$ are the AdS radius and Newton's constant, respectively) 
\be
S={l\over 16\pi G}\int dxdtdp {\sqrt g}(R+2/l^2)=0
\ee 
coincide with the equations describing two decoupled  
Chern-Simons (CS) models with the combined action \cite{sh}
\be
S={l\over 16\pi G}
Tr\int dxdpdt \epsilon^{\mu\nu\lambda}({\hat A}_{\mu}^{\pm}\partial_{\nu} {\hat A}_{\lambda}^{\pm} + 
{\hat A}_{\mu}^{\pm}{\hat A}_{\nu}^{\pm}{\hat A}_{\lambda}^{\pm})
\ee 
The chiral connections ${\hat A}_{\mu}^{\pm}$ are matrices that can be expanded  
in the basis spanned by the generators ${\hat L}^{\pm}_{0,\pm 1}$ of the   
algebra  
$SL(2,R)\times SL(2,R)=SO(2,2)$. They obey the commutation relations
$
[{\hat L}^{\pm}_n,{\hat L}^{\pm}_m]=(n-m){\hat L}^{\pm}_{n+m}
$
and are normalized,    
$Tr {\hat L}^{\pm}_{n}{\hat L}^{\pm}_m= {1\over 2}\delta_{n0}\delta_{m0}
-\delta_{n1}\delta_{m,-1}$.

The topological action (33) then 
reduces to a pure boundary term while the equation of motion becomes 
that of null curvature
\be
\partial_{\mu}{\hat A}_{\nu}^{\pm}+{\hat A}_{\mu}^{\pm}{\hat A}_{\nu}^{\pm}-(\mu\leftrightarrow\nu)=0
\ee
Parameterizing its solutions in terms of an arbitrary group element ${\hat \chi}_{\pm}$ and functions $p_{\pm}(x_{\pm})$ and $\mu_{\pm}(x_{\pm})$ 
\be
{\hat A}^{\pm}(x,p,t)={\hat \chi}^{-1}_{\pm}(p)
{\hat L}_0(\mu_{\pm}\pm p_{\pm}dx){\hat \chi}_{\pm}(p)
\ee
 one finds this equation to be equivalent to 
\be 
{\dot p}_{\pm}\mp \mu_{\pm}^{\prime}=0
\ee
which, in turn, coincides with one of the above solvable equations, 
provided that the chemical potentials $\mu_{\pm}$ conjugate to the variables $p_{\pm}$  
are given by the derivatives 
\be
\mu_{\pm}={\delta H^{\pm}\over \delta p_{\pm}}
\ee
Choosing the Hamiltonian appropriately one can then reproduce 
the solvable (m)KdV, Gardner, and other equations. 
In particular, the KdV family is recovered for  
\bea
{\hat A}_p^{\pm}(x,p,t)={1\over p}{\hat L}^{\pm}_0~~~~~~
{\hat A}_x^{\pm}(x,p,t)= p{\hat L}^{\pm}_1-{p_{\pm}\over p}{\hat L}^{\pm}_{-1}~~~~~~~~
\nonumber\\
{\hat A}_t^{\pm}(x,p,t)= {p\mu_{\pm}}{\hat L}^{\pm}_1-\mu_{\pm}^{\prime}{\hat L}^{\pm}_{0}
+{{\mu_{\pm}^{\prime\prime}-\mu_{\pm} p_{\pm}}\over 2p}{\hat L}^{\pm}_{-1}~~~~~~~~~~
\eea
which expressions are manifestly Lorentz-non-invariant for all $k>1$.

Moreover, the equation of motion (34) can be converted into that of the gravity 
model (32) under the identification of the $3d$ metric 
\be
g_{\mu\nu}={l^2\over 4}<(A_{\mu}^{+}-A_{\mu}^{-})(A_{\nu}^{+}-A_{\nu}^{-})>
\ee
On the gravity side the  
different saddle points of the coherent states path integral can be identified 
as globally distinguishable (but locally $AdS_3$) classical solutions. 
In particular, it can be shown that the only minima of the action (32) 
corresponding to the boundary Hamiltonian $H_{1}+H_{2}$
 are those with constant (negative) curvature.
The two competing minima are the thermal $AdS_3$ and BTZ (Banados-Teitelboim- Zanelli)
black hole. 

However, by introducing higher order terms $H_k$ with $k\geq 3$ 
one can generate new KdV-charged black hole configurations \cite{sh}. 
The corresponding boundary theory is encoded in 
the boundary conditions for the connection (35), by varying which one can explore  
a variety of the integrable $1d$ systems. 

The standard LL with $k=1$ is reproduced by introducing 
the original Brown-Henneaux boundary conditions
with constant $\mu_{\pm}\sim p_{\pm}$, the outer/inner horizons
being located at $p_{>/<}=(p_+\pm p_-)/2$. The dual metric 
\bea
ds^2={dp^2\over p^2}
+(p^2-2(p_{+}+p_{-})+{p^2_{+}+p^2_{-}\over p^2}) dt^2+~~~~~\nonumber\\
(p^2+2(p_{+}+p_{-})+{p^2_{+}+p^2_{-}\over p^2}) dx^2+(p_{+}p_{-})dx dt~~~~~
\eea
describes a rotating BTZ black hole 
with the event horizon but no curvature singularity.

For static, yet non-constant, $p_{\pm}(x)$ the 
corresponding boundary solutions possess non-trivial global charges given by the 
chiral surface integrals $H_{k}^{\pm}$ while their bulk counterparts   
can be regarded as black holes with multi-graviton excitations ('soft hair') \cite{sh}.

The general solution can be obtained by acting on
the ground state (e.g., BTZ black hole) with elements of the asymptotic symmetry
group commuting with the Hamiltonian. This way one can construct various constant curvature, yet 
locally AdS, spacetimes with anisotropic Lifshitz scaling and 
dynamical exponent $z = 2k -1$. This opens up the possibility of studying nonrelativistic
holography without the need of bulk geometries which are asymptotically Lifshitz spacetimes. 
Furthermore, in the case of a higher-spin symmetry $SL(M,R)$ the list of attainable gravitational backgrounds may include the asymptotically 
Lobachevsky, Schroedinger, warped $AdS$, etc. spacetimes \cite{carol}.

Shocks and other abrupt perturbations are characterized by 
FS breakdowns and emergence of folds where
the spatial derivative $\rho^{\prime}$ diverges,
thereby requiring several pairs of the bounding momenta $p_{\pm}$.
In the presence of shocks the conventional spacetime 
hydrodynamics becomes insufficient for describing long-time behavior,
although the full-fledged phase space hydrodynamics can avoid such problems.

In that regard, particularly interesting are the non-stationary 
configurations representing particles 
released from a confining potential which gets suddenly switched on/off  \cite{quench}.
Such quenching profiles generically have spacelike boundaries where 
the saddle point solutions of the collective field hydrodynamics
diverge at finite times and the semi-classical description fails.
Ascertaining the emergent spacetimes and their dynamics then requires a detailed 
study of fluctuations around the pertinent saddle points.
\\ 
 
{\it Reductions and generalizations}
\\

Despite having been repeatedly stated 
and extensively analyzed at the level of salient 
symmetries and concomitant algebraic properties,
the general gravity/fluid correspondence 
in dimensions $d>1$ has not yet been put to much of a practical use \cite{fluid}. 

Specifically, such  a relationship was shown to exist 
between the solutions of classical $d+2$-dimensional gravity  
and their $d+1$-dimensional hydrodynamic counterparts, whereby 
the former would be given by the metric 
\be
ds^2={dp^2\over f(p)p^2}+
p^2(\Delta_{\mu\nu}-f(p)u_{\mu}u_{\nu})dx^{\mu}dx^{\nu}
\ee
parameterized in terms of the 
spacetime-dependent covariant velocity $u_{\mu}({\bf x},t)$ and local 
temperature $T({\bf x},t)$ \cite{fluid}. The latter satisfy the hydrodynamic equations
on a fixed background, provided that  
$f(p)=1-(4\pi T/p)^d$ and $\Delta_{\mu\nu}=g_{\mu\nu}+u_{\mu}u_{\nu}$.
Thus, a given fluid profile can be associated with a certain asymptotically $AdS_{d+2}$-like
spacetime with a horizon located at $p_h=4\pi T$. 

A still more general (asymptotically accurate) 
solution can be constructed with the use of the metric ansatz 
\be
g_{\mu\nu}({\bf x},p,t)={g^{(0)}_{\mu\nu}\over p^2}+
g^{(2)}_{\mu\nu}
+{p^2\over 4}g^{(2)}g^{(0)}g^{(2)}_{\mu\nu} 
\ee
with arbitrary functions $g^{(0,2)}_{\mu\nu}({\bf x},t)$ \cite{fluid}.

In contrast to the generic case of $d>1$, 
pure gravity  in $d\leq 1$ does not support any finite energy excitations. 
Therefore, the  fluid/gravity correspondence takes on a particularly simple form where  
the dual bulk theory appears to be non-dynamical and fully determined 
by the gapless boundary degrees of freedom ('boundary gravitons').

Of course, such a scenario of 'holography light'  
does not quite rise to the level of genuine holographic correspondence
where the bulk theory is supposed to have a non-trivial quantum dynamics 
which gets quenched and turns classical only in a certain (large $N$) limit.  
It should be noted, however, that, barring a few exceptions, the customarily assumed  
'classicality' of the bulk geometry (regardless of whether or not the $1/N$- and/or 'stringy' corrections are important) and a complete neglect of any back-reaction on the fixed
background metric appear to be by far 
the most common approximations routinely made in the absolute majority 
of all the $AdS/CMT$ calculations performed so far \cite{ads,holo}. 

Nonetheless, there are still important differences between the situations in $d=0$ and $d=1$.
As per the above discussion, the latter is described by the LL action of two 
chiral $1d$ bosons $\phi_{\pm}(x_{\pm})=\ln\Phi^{\prime}_{\pm}(x_{\pm})$ 
\be
S^{\pm}_{LL}=\int dx dt {\Phi^{\prime\prime}_{\pm}
({\dot \Phi}^{\prime}_{\pm}\mp\Phi^{\prime\prime}_{\pm})\over (\Phi^{\prime}_{\pm})^2}=
\int dx dt\phi^{\prime}_{\pm}({\dot \phi}_{\pm}\mp\phi^{\prime}_{\pm})
\ee
This action can also be obtained from the more general 
Alekseev-Shatashvili functional which  
performs path-integral quantization on the co-adjoint orbit of the (double) Virasoro group. Alternatively, it can be identified with the large central charge limit
in the conformal Liouville model, thus relating the latter to its namesake (Liouville) 
theorem governing the phase space dynamics in the (semi)classical limit.

Besides, this action can be viewed as a complexity functional 
 defining an associated quantum-information type of geometry on the Virasoro group,
its lower bound being given by the length of a proper geodesic on the co-adjoint orbit \cite{qc}.

In turn, the extensively studied case of $d=0$ 
can be attained in the $AdS_3$ theory by taking the 
limit of a vanishing length of the compactified spatial dimension.
The resulting $AdS_2$ bulk theory, as well as its JT (Jackiw-Teitelboim) 
extension, support a pseudo-Goldstone 
time reparametrization mode with the $1d$ boundary action given by the Schwarzian derivative
\cite{syk}. 

Equivalently, it can be cast in terms of the Liouville quantum mechanics on the quotient $Diff(S^1)/PSL(2,R)$ with the action  
\be
S_{L}=\int dt ({1\over 2}{\dot \phi}^{2}+\lambda e^{\phi})
\ee
for $\phi(t)=\ln{\dot \Phi}(t)$ where 
$t\to \Phi(t)$ is a diffeomorphism of the thermal circle.  In the context of 
the space-less random SYK (Sachdev-Ye-Kitaev) and non-random tensor models
this orbit emerges as the result of factoring out the subspace of 
zero modes reflective of the $SL(2,R)$ symmetry of the conformal saddle-point  
solutions \cite{syk}. The integrable $1d$ dynamics in such models is spatially ultra-local 
and corresponds to $z=\infty$, thus being reminiscent of the popular $AdS/CMT$ schemes \cite{ads}. 

Notably, in contrast to the marginal nature of the $1d$ LL theory where the interaction remains 
important at all energy/temperature scales, in the SYK/tensor models it is strongly relevant 
in the infrared, thus only affecting the conformal mean-field 
solutions below a certain energy/temperature scale.   
Also, the maximally chaotic $AdS_2/JT$ gravity can be dual not to a certain quantum 
mechanical ($d=0$) theory but (as in the case of SYK) a random ensemble thereof.
For comparison, in $d=1$ neither the boundary theory (43) saturates the chaos bound, nor 
is the bulk behavior dominated by pure gravity. 

In practice, establishing the SYK-to-$AdS_2/JT$ duality involves matching thermodynamic properties of the two systems, alongside their various correlation functions.  
However, achieving this correspondence beyond the lowest order 
(two-point) correlation functions requires one to introduce additional 
'matter' fields in the bulk which represents a tower of higher-spin operators 
with the anomalous dimensions that all scale comparably with $1/l$ \cite{syk}. 

Likewise, in the KdV-to-$AdS_3$ correspondence the entropy, 
free energy, etc. can be matched as well, giving rise to the dependencies 
$S={\pi\over 4}\sum_{\pm}p_{\pm}^d\sim T^{1/z}$ and $E=\sum_{\pm}<{\hat H}^{\pm}_k>\sim T^{1+1/z} $, provided that one chooses 
$\mu_{\pm}\sim T$ in order for the metrics (39) to remain regular everywhere in space. 
Notably, the KdV-charged black holes' thermodynamics differs from that 
of the usual BTZ ones \cite{sh}.

Comparison between the pertinent microstates on both sides of the latter correspondence 
relies on the fact that the $2d$ phase space can be spanned by 
the overcomplete basis of coherent states 
$
|\Psi>=\exp(i\sum_{nm} c_{nm}{\hat W}_{nm})|0>
$
while the boundary $1d$ theory operates in the Hilbert space spanned by the vectors
$
|\pm n>=\prod_n {\hat p}_{\pm}^n|0>
$.
Employing this basis the correlators of a bulk field $\hat O$
of mass $m$ and dimension $\Delta=(d+1)/2\pm lm$ can be evaluated by the 
saddle point method, thereby resulting 
in the semiclassical expression for the (real-time) two-point function 
\be
{G}_{OO}(x,p,t)\sim \exp(-\Delta\int[g_{pp}dp^2-g_{tt}dt^2+g_{xx}dx^2]^{1/2})
\ee
where the line integral is taken over the $3d$ geodesic connecting the end points. 

Placing the end points of this correlation function
on the boundary yields the single-particle boundary
propagator. Fourier transforming this expression in the spacetime domain one then obtains 
\be
G_{OO}(\omega,k)\sim \exp(-\int dp[g_{pp}({\Delta^2\over l^2}-{\omega^2\over g_{tt}}+{k^2\over g_{xx}})]^{1/2})
\ee
For instance, the BTZ bulk metric (40) yields the following propagator of massless  
$3d$ bulk fermions with the dimension $\Delta_{\Psi}=1$ and spin $1/2$ \cite{LL}
\be
G^{\pm}_{\Psi\Psi}(\omega,k)=({\omega\pm k\over \omega\mp k})^{1/2}
\ee
which gives rise to the power-law spacetime behavior of the boundary propagator
$G^{\pm}_{\Psi\Psi}(t,x)\sim 1/|x_{\pm}|^2$.
These results should of course be distinguished from the standard LL propagator 
$G^{\pm}_{\psi\psi}(\omega,k)=1/({\omega\mp k})$ of free chiral fermions with the dimension 
 $\Delta_{\psi}=1/2$.

In contrast, using the metric build out of the solitonic solution (29)
changes the spacetime decay from algebraic to exponential, 
$G^{\pm}_{\Psi\Psi}(t,x)\sim \exp(-\Delta |x_{\pm}|/l)$.

In turn, the two-particle (energy) excitations 
representing gapless boundary gravitons 
remain propagating, thus featuring the ordinary ballistic pole 
\be
G_{\epsilon\epsilon}(\omega,k)=
\int dtdxe^{i(kx-\omega t)}[G^{\pm}_{\psi\psi}(x,t)]^2
\sim
{k^2\over k^2-\omega^2}
\ee
Alternatively, this energy correlation function can be obtained from the correlator  
$<w(x,p,t)w(x^{\prime},p^{\prime},t^{\prime})>$ 
computed as the path integral over the Wigner function. 

The ballistic behavior (48) should be contrasted against the diffusive one observed 
in, e.g., a chain of coupled SYK models, $G_{\epsilon\epsilon}(\omega,k)
\sim {k^2/(k^2+i\omega)}$, which would be indicative of a (maximally) 
chaotic state \cite{syk}. 

Further possible generalizations of the collective field hydrodynamics 
include incorporation of the momentum Berry curvature in Eq.(6),  
exploration of the effects of viscosity, 
generic dispersion with $z\neq 1,2$, etc. It would also be interesting to 
investigate a development of turbulence which has long been known to 
harbor some important connections to quantum gravity.
\\

{\it Discussion}
\\

In this note the Wigner function representation of generic 
quantum systems was discussed from the standpoint of pinpointing 
the possible origin of the hypothetical generalized holographic correspondence. 
To that end, using the Kirillov-Kostant 
procedure of phase space quantization via the coherent state path integral
and collective field hydrodynamics may seem rather promising. 

Specifically, in line with the holographic lore, path integral quantization on a  
co-adjoint orbit of the $W_{\infty}$ group of the volume-preserving diffeomorphisms
of the phase space exposes an intrinsic relationship 
between the $2d+1$-dimensional 'bulk' description 
and the $d+1$-dimensional 'boundary' hydrodynamics. 
The quantum bulk dynamics is described by the action composed of the 
$W_{\infty}$ generators while the corresponding boundary 
variables are given by the moments (17) of the Wigner function.
Systematically implementing this program can then be thought 
of as 'deriving' the sought-out holographic duality.

Importantly, such a generic form of correspondence neither requires 
a reference to some underlying 'string' theory, nor 
does it impose any particular symmetry conditions on 
the boundary system in question, while the putative 
bulk description does not necessarily have a gravity-like appearance.

Nonetheless, in the case of $d=1$ the corresponding $3d$ bulk behavior can indeed be cast in terms of the (doubled) Chern-Simons theory or, equivalently, 
non-dynamical Einstein gravity with a negative cosmological constant. 
Furthermore, if the boundary Hamiltonian belongs to the 
integrable (e.g., KdV) family, the corresponding set of the bulk metrics may include the familiar BTZ, as would be generally anticipated in line
with the $AdS/CFT$ paradigm \cite{ads}. 

Furthermore, in $d=1$ the phase space description can be viewed as 
a formally exact (non-linear) bosonization of the boundary system. 
Many of such systems appear to be integrable (hence, non-ergodic) and possess an infinite number of locally conserved currents given by the various moments of the Wigner function which obey the equations of zero-entropy generalized quantum hydrodynamics of the GGE type.

In higher dimensions the Wigner function's moments, too, serve as coefficients in the series expansions of the would-be local bulk metrics over 
the powers of the momentum $\bf p$. 
Although with the increasing spatial dimension 
the hydrodynamic description becomes less accurate, it remains
capable of capturing the salient features of the quantum phase space dynamics
governed by the conservation laws.  

To summarize, the use of the phase space approach brings out the 
intrinsic correspondence between a formally exact $2d+1$-dimensional and a less accurate ('coarse-grained')  
$d+1$-dimensional hydrodynamic descriptions of a given quantum system. 
In this general setting, neither the latter needs to be a conformal field 
theory, nor does the former have to have the appearance of gravity. 

In those $d=0$ and $d=1$ cases where the bulk indeed appears to be  
amenable to a gravitational description
the gravity theory has no dynamics of its own and is
fully determined by the boundary degrees of freedom. 
Accordingly, the viable bulk metrics can be 
mapped onto the solutions of the boundary 
hydrodynamics. 

In that regard, the holographic custom of picking out a particular 
metric and claiming some sort of the  
Einstein-Maxwell-dilaton theory to be the proper bulk dual of a
certain strongly correlated system does not appear to be backed by the above conclusions. 
Nevertheless, in some cases, including $d=0$ and $d=1$, 
certain phenomenological predictions
may indeed turn out to be right - albeit, quite possibly, for the wrong reason.  
On the other hand, if the essentials of practical holography were to amount to little else but  
hydrodynamics then the whole issue would become largely moot and void.

\end{document}